\documentclass[preprint,12pt,sort&compress]{elsarticle}
\journal{Physica A}
\usepackage{amsmath,amssymb,graphicx,color,lipsum,mathrsfs}
\newcommand{\ie}{\emph{i.e.}}
\newcommand{\eref}[1]{Eq.~(\ref{#1})}
\begin{document}
\begin{frontmatter}
\title{Analysis of ground state in random bipartite matching}
\author[fr]{Gui-Yuan Shi}
\author[fr]{Yi-Xiu Kong}
\author[fr]{Hao Liao\corref{cor}}
\ead{hao.liao@unifr.ch}
\author[fr]{Yi-Cheng Zhang}
\ead{yi-cheng.zhang@unifr.ch}
\cortext[cor]{Corresponding author}
\address[fr]{Physics Department, University of Fribourg, Chemin du Mus\'{e}e 3, CH-1700 Fribourg, Switzerland}

\begin{abstract}
In human society, a lot of social phenomena can be concluded into a mathematical problem called the bipartite matching, one of the most well known model is the marriage problem proposed by Gale and Shapley. In this article, we try to find out some intrinsic properties of the ground state of this model and thus gain more insights and ideas about the matching problem. We apply Kuhn-Munkres Algorithm to find out the numerical ground state solution of the system. The simulation result proves the previous theoretical analysis using replica method. In the result, we also find out the amount of blocking pairs which can be regarded as a representative of the system stability. Furthermore, we discover that the connectivity in the bipartite matching problem has a great impact on the stability of the ground state, and the system will become more unstable if there were more connections between men and women.
\end{abstract}

\end{frontmatter}

\section{Introduction}
Bipartite matching problem, which is to match disjoint two groups of agents, is widely seen in human society, for instance, the marriage problem between men and women, college admission problem between students and universities, assignment between workers and jobs, and also the choice making between buyers and sellers.

Gale and Shapley first introduced the stable marriage problem, a one-to-one two side matching~\cite{marriage}, which is still one of the most important bipartite matching problems. Gale-Shapley Algorithm guarantees the existence of a stable solution that is also the optimal stable solution for the active side. Due to its multiple applications in society and its intriguing fascinating properties, not only economists but also the statistic physicists are attracted by the bipartite matching problem. By giving up the restriction to stabilize the system, Parisi and his colleagues used the replica method of spin glass theory to study the global optimal solution of one-to-one two side matching problem~\cite{replica,parisi}. Later, Zhang et al. studied the scaling behavior of Gale-Shapley model, partial information matching and analyzed the common features of all the stable solutions~\cite{scaling,partial,happier}. Moreover, Dzierzawa introduced the acceptance threshold and thus improved the matching result of the passive side~\cite{statistics}. Recently, Zhou et al studied the bidirectional selection problem, provided us a new approach from human society network~\cite{twoway,network}.

This article is organized in the following way. First we introduce some fundamental background about Gale-Shapley model. Then we use numerical simulations to study the optimal solution based on the bipartite marriage problem. By bringing the Kuhn-Munkres Algorithm~\cite{algorithm,hungarian} into the marriage problem, we capture the ground state of the system. Next we analyze the average energy and other properties of this solution in detail. After that, the number of blocking pairs in the ground state proposed by Zhang~\cite{scaling} is given by the simulation result, the dynamics is also discussed here. At last, we make a simple a quantity analysis of the stability of system ground state.

\section{Model}
Here we study the optimization solution of the bipartite matching problem on the basis of the Gale and Shapley marriage model. This model consists of two sets of agent,  $\mathscr{M} = \{ m_1, m_2,\ldots, m_n\}$ standing for n men and $\mathscr{W}= \{w_1, w_2,\ldots, w_n\}$ for n women. A final outcome of the marriage problem is a one-to-one matching of men and women, \ie \ an invertible bijection $x:\mathscr{M} \to \mathscr{W}$. An outcome $x$ can be denoted as:
$$x=[(m_1,x(m_1)),(m_2,x(m_2)),\ldots,(m_n,x(m_n))],$$
which $x(m_i)=w_{\alpha}$ means the woman who matched with man $m_i$, and $x^{-1}(w_{\alpha})=m_i$ is the man matched with $w_\alpha$~\cite{roth}.

One of the most attractive questions all along these years is the optimal solution of the system. In order to find out the optimal solution, we need a measurement for the satisfaction of each man and woman. Compared with the way of nature that it always looks for the ground state, we also assign an energy term $\varepsilon_i$ to agent $i$, to represent how satisfactory he/she is. The energy term actually represents the ranking of the assigned mate for agent $i$. The smaller $\varepsilon$, the happier he/she is. Several previous works~\cite{replica,parisi,scaling,happier,partial,statistics} postulated the energy term $\varepsilon$ follows a discrete uniform distribution like $\varepsilon=1, 2, \ldots, n$ in their model, but we think this might be extended to some other distributions. Since we have a large population $2n$ in the system, there would be no much difference between the expectation values of discrete distribution and the continuous distribution. So here we suggest a more soften assumption that the energy $\varepsilon$ is randomly uniform distributed on $[0, 1]$, and we predict that it will not change the result much from the previous work~\cite{replica,parisi,scaling}. With the description of satisfaction of each agent, the optimal solution problem will be rephrased into a minimum energy problem.

For the convenience, we denote the men with Latin alphabet, and women with Greek alphabet. We define matrix $M$ and $W$, in which $M_{i,\alpha}$  is the energy of man $m_i$ if women $w_\alpha$ was matched to him, all the same, $W_{\beta,j}$ is the energy of woman $w_{\beta}$ if man $m_j$ was matched to her. For a given assignment $x=[(m_1,x(m_1)),(m_2,x(m_2)),\ldots,(m_n,x(m_n))]$, we can calculate the average energy of each man and woman:

\begin{equation}\label{define1}
\begin{split}
\varepsilon_M(x)=\frac{1}{n}\sum_{i=1}^{n}M_{i,x(m_i)},
\end{split}
\end{equation}

\begin{equation}\label{define2}
\begin{split}
\varepsilon_W(x)=\frac{1}{n}\sum_{\alpha=1}^{n}W_{\alpha,x^{-1}(w_{\alpha})}=\frac{1}{n}\sum_{i=1}^{n}W_{x(m_i),i}.
\end{split}
\end{equation}
Then we have the average energy per person:

\begin{equation}\label{define3}
\begin{split}
\varepsilon_H(x)=\frac{1}{2n}[\sum_{i=1}^{n}M_{i,x(m_i)}+\sum_{i=1}^{n}W_{x(m_i),i}]
                =\frac{1}{n}\sum_{i=1}^{n}H_{i,x(m_i)}.
\end{split}
\end{equation}
Here we define the matrix $H=\frac{M+W^T}{2}$, where the element $H_{i,\alpha}$ is the mean energy of the man $m_i$ and woman $w_\alpha$. With the matrix $H$, we now are able to solve the optimal solution $\varepsilon_H^{min}$ \ie \ the minimum of $\varepsilon_H$, and the corresponding matching $x$ is called the ground state.

In the example of a matching problem which consists $n$ men and $n$ women, there will be $n!$ different states of the system in total. Obviously it's neither possible nor clever to find out the ground state by Exhaustive method. Facing with a similar situation, Kuhn~\cite{hungarian} and Munkres~\cite{algorithm} developed an algorithm called $Kuhn-Munkres$ $algorithm$, to solve minimum solution in a weighted bipartite matching. We think this method can also be applied to our problem here. The total energy matrix $H$ can be regarded as the cost matrix in the bipartite assignment problem. In this way, it is possible to dig out the matching corresponding to the ground state which has the lowest global energy only spending time of $O(n^4)$. But with previous study of the matching problem, it's unfortunate that the ground state of this matching problem is not stationary state, which means it is not stable. Here we call a matching $x$ is stable only if there was no pairs of man and woman who prefer each other than their assigned mate in $x$, that is to say, for each individual there is no any better choice to improve their personal situation. But if such pair of man $m_i$ and woman $w_\alpha$ exists, we name it a blocking pair~\cite{scaling,path}, abbreviated as $BP$ hereafter. The more $BPs$ a system contains, the more unstable it is. We are curious about the quantity of $BPs$ in a matching problem, because it can be treated as a representative for the stability of a certain matching~\cite{scaling}.

\section{K-M Algorithm}
Before we give a description of this algorithm~\cite{algorithm,hungarian}, we'd like to introduce two lemmas:

\textbf{Lemma 1.} The equivalent statement of K{\"o}nig's theorem: in matrix $H$, the biggest amount of independent zero elements we can find is equal to the minimum number of lines (rows or columns) that covers all the zero elements.

\textbf{Lemma 2.} Obviously, the minimal assignment solution does not change, if we add or subtract the same number in any rows or columns at the same time.

\fbox{
  \begin{minipage}{.91\textwidth}

\textbf{Algorithm description}

\textbf{Step 1.} Starting with a given matrix $H$, for each row and column, we subtract its minimum value from the rows and columns (thus we create one or several zeros in those rows and columns). We denote the new matrix $H_1$.

\textbf{Step 2.} Find the smallest set of lines $L_k$(vertical and horizontal), which consists of $n_k$ lines, to cover all the zeros in the matrix $H_1$.

\textbf{Step 3.} If $n_k=n$, then we will have $n$ independent zeros, the corresponding $n$ positions represent the final assignment.

\textbf{Step 4.} If $n_k<n$, find out the smallest matrix element $a_k$ which was not covered by any of $L_k$. We add $a_k$  to all the uncovered rows and subtract $a_k$ from all the uncovered columns. Then return to step 2.
  \end{minipage}
}

\section{Simulation and Analysis}
We applied this Kuhn-Munkres Algorithm to our model, and studied the optimal solution \ie ground state of the marriage problem where the population of men or women $n=100, 200, 300, \ldots, 1000$.

With each fixed population of $n$ men and $n$ women, we did $100$ trials starting with randomly generated $H$ matrix, the result of the average energy per person in the ground state $\langle\varepsilon_H^{min}\rangle$ and its standard deviation $\sigma$ are shown in Tab.1. We see that the standard deviation can be estimited as the reciprocal of n. That is  because each agent is identical to another who are weakly correlated, by the Central Limit Theorem~\cite{clt}, we can know the $\varepsilon_H^{min}$ follows the normal distribution. Furthermore, we make 10000 trials when n=100, the simulation result proves our analysis, shown in Fig.1.

\begin{table}[htb]\label{datasets}\scriptsize
\caption{Mean and Standard Deviation of Energy Per Person}
\begin{center}
\begin{tabular}{ccccccccccc}
  \hline \hline
   n      &100        &200         &300      &400    &500   &600 &700 &800 &900  \\ \hline
   $\langle\varepsilon_H^{min}\rangle$      &0.0802        &0.0569       &0.0464   &0.0405  &0.0361  &0.0329 &0.0305 &0.0285 &0.0269  \\
   $\sigma$        &0.0034  &0.0017  &0.0011 &0.00091 &0.00070   &0.00056 &0.00047 &0.00042 &0.00038   \\
   \hline \hline
\end{tabular}
\end{center}
\end{table}

\begin{figure}[thb]
\center\scalebox{0.3}[0.3]{\rotatebox{0}{\includegraphics{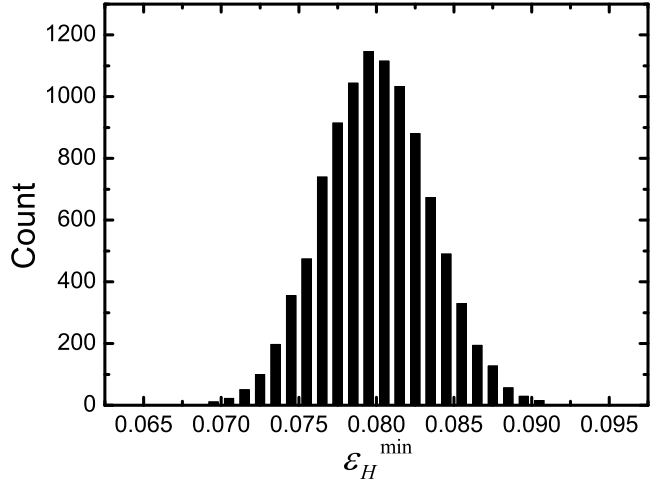}}}
\caption{The distribution of 10000 trials on $\varepsilon_H^{min}$ when n = 100.}\label{Fig0}
\end{figure}

The $\langle\varepsilon_H^{min}\rangle$ for the $100$ trials is shown in Fig.2. With a power exponent fit, we get the power of $-0.495$ and also the expectation of average energy per person is $\frac{0.808}{\sqrt{n}}$, which is very close to the predicted value $\frac{0.8085}{\sqrt{n}}$ of Zhang et al~\cite{scaling} with replica method.

\begin{figure}[htb]
\center\scalebox{0.3}[0.3]{\rotatebox{0}{\includegraphics{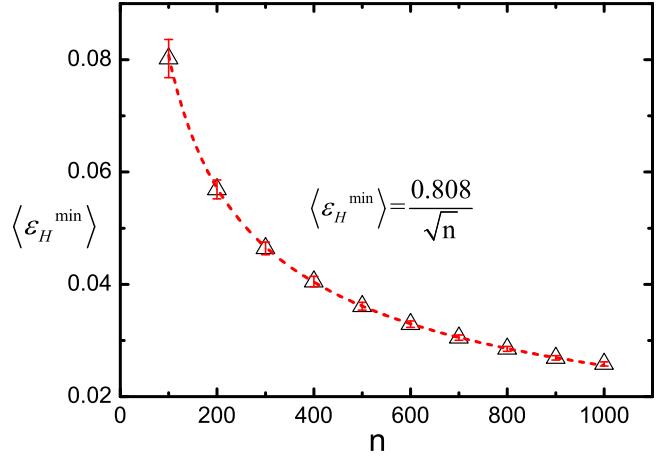}}}
\caption{The relationship between $\langle\varepsilon_H^{min}\rangle$ and the system scale $n$, for $n= 100,200,300,\ldots,1000$. The dashed line indicate the fitting of $\frac{0.808}{\sqrt{n}}$.}\label{Fig1}
\end{figure}

Furthermore, the distribution of personal energy $\varepsilon$ is studied here as well, shown in Fig.2. By taking the semi-logarithmic, we find that the probability distribution function $f(\varepsilon)$ decays faster than exponent, but it can be fitted almost perfectly by an exponent function with its index being quadratic function:

\begin{figure}[htb]
\center\scalebox{0.3}[0.3]{\rotatebox{0}{\includegraphics{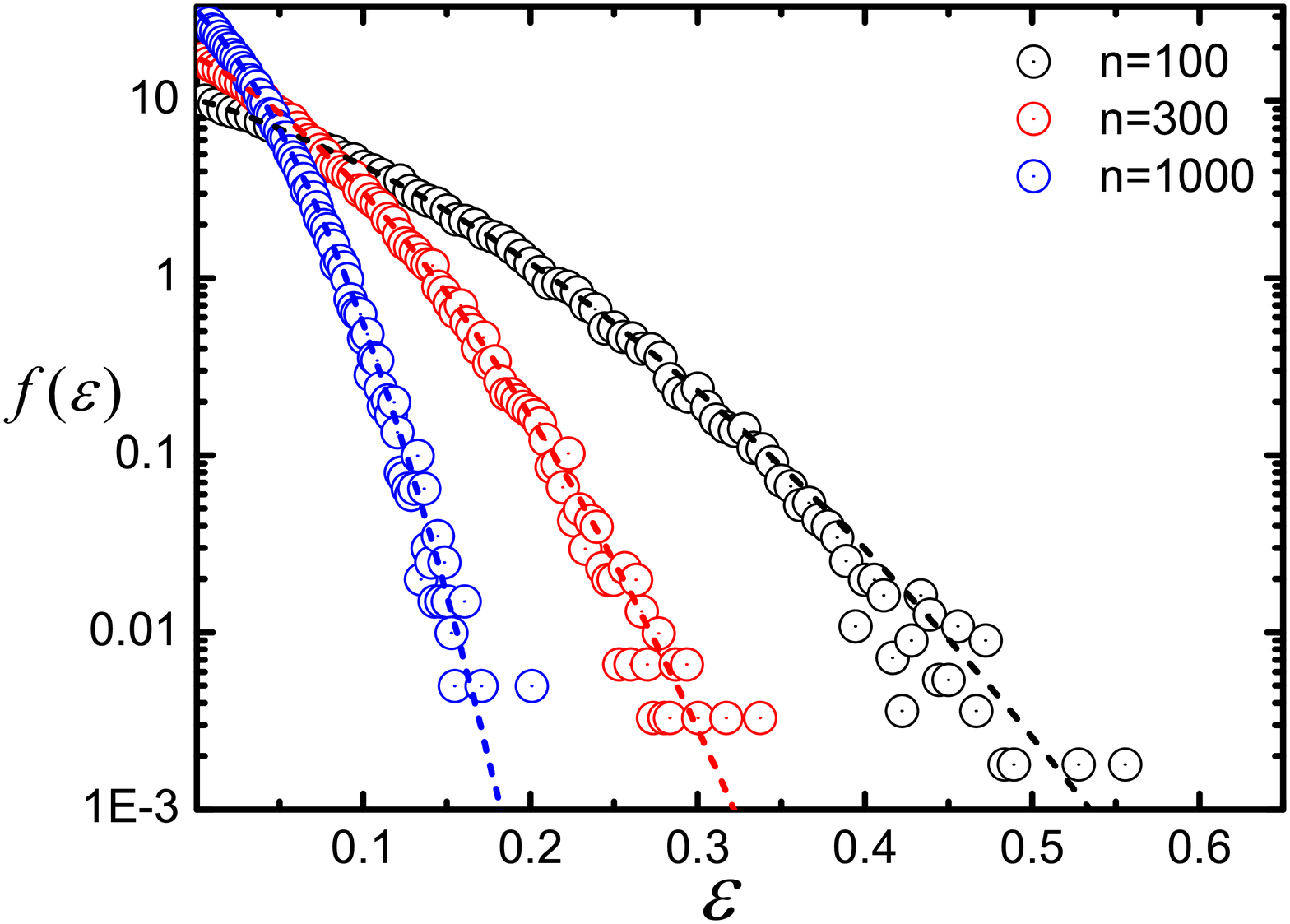}}}
\caption{(Color online) The PDF of $\varepsilon$ with $n=100$(black circles), $300$(red circles), $1000$(blue circles). The dashed lines are the corresponding fitting by~\eref{fit2}.}\label{Fig2}
\end{figure}

\begin{equation}\label{fit1}
\begin{split}
\frac{f(\varepsilon)}{\sqrt{n}}=e^{-a(\sqrt{n}\varepsilon)^{2}-b(\sqrt{n}\varepsilon)+c}.
\end{split}
\end{equation}
If we treat the $\sqrt{n}\varepsilon$ as the argument and the $\frac{f(\varepsilon)}{\sqrt{n}}$ as the dependent variable (in this way we normalized the scaling factor), and then fit the parameter $a$, $b$ and $c$ in~\eref{fit1} with the simulation results, we get $a=0.19$, $b=0.68$, and $c=0.02$, so the distribution function can be written as:
\begin{figure}[htb]
\center\scalebox{0.3}[0.3]{\rotatebox{0}{\includegraphics{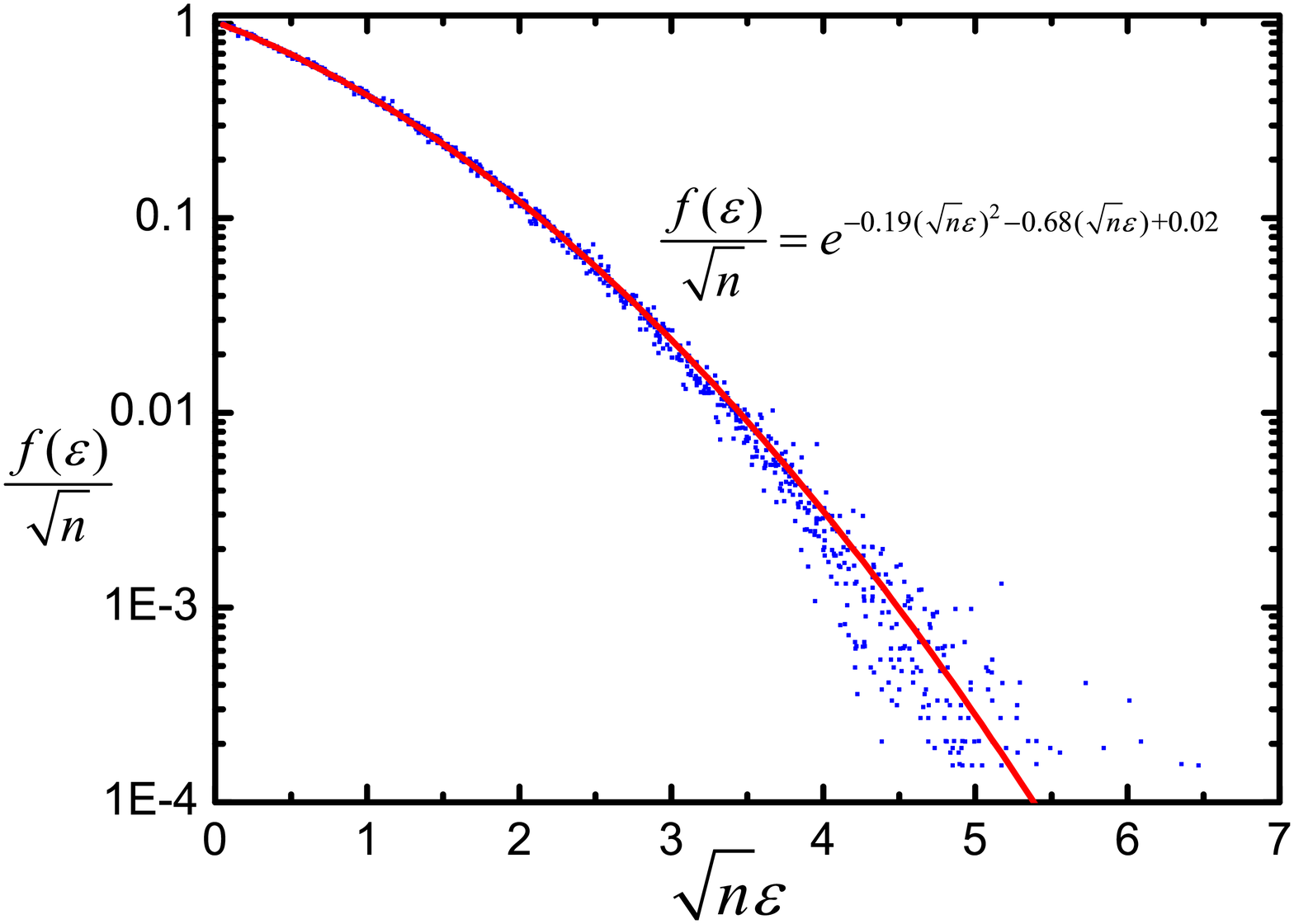}}}
\caption{The fitting of $\frac{f(\varepsilon)}{\sqrt{n}}$ against $\sqrt{n}\varepsilon$.  }\label{Fig3}
\end{figure}

\begin{equation}\label{fit2}
\begin{split}
\frac{f(\varepsilon)}{\sqrt{n}}=e^{-0.19(\sqrt{n}\varepsilon)^{2}-0.68(\sqrt{n}\varepsilon)+0.02}.
\end{split}
\end{equation}
To ensure that these parameters are reasonable, we integrate the distribution function and it shows that:

\begin{equation}\label{check1}
\begin{split}
\int_{0}^{1}f(\varepsilon)\,d\varepsilon=1.03,
\end{split}
\end{equation}

\begin{equation}\label{check2}
\begin{split}
\int_{0}^{1}\varepsilon f(\varepsilon)\,d\varepsilon=\frac{0.84}{\sqrt{n}}.
\end{split}
\end{equation}
The result is a little bigger than the actual value, which means the actual distribution function decays even faster than $e^{{-\varepsilon}^2}$, but this function is still in good accordance with the simulation result.

\section{Blocking Pairs}

As Zhang et al mentioned~\cite{scaling}, the quantity of blocking pairs can be used as a measurement of the stability of a solution, and he gave a rough estimation of the number of blocking pairs being $(n\varepsilon)^2 \sim 0.65n$ .

Now we try to think more about this. First let us consider the possibility that a man cannot find a $BP$. Imagine a man $i$ with energy $\varepsilon$, there would be $n\varepsilon$ women who are possible to form a $BP$ with man $i$, \ie $F_{i,\alpha_1}, F_{i,\alpha_2},\ldots,F_{i,\alpha_{n\varepsilon}}<F_{i,x(i)}$.  If these $n\varepsilon$ women satisfied the following situation that their current mate were all better than man $i$, which can be represented as:
$$G_{\alpha_1,x^{-1}(\alpha_1)}<G_{\alpha_1,i}, G_{\alpha_2,x^{-1}(\alpha_2)}<G_{\alpha_2,i},\ldots, G_{\alpha_{n\varepsilon},x^{-1}(\alpha_{n\varepsilon})}<G_{\alpha_{n\varepsilon},i},$$
there would not be any blocking pair forming with man $i$.

In the process of chasing ground state, if we neglect other influencing factors and fix the $G_{\alpha,i}$, we can see that the possibility of woman $\alpha$ matching to man $i$ gets bigger when $F_{i,\alpha}$ gets smaller. So in the situation that woman $\alpha$ was not matched to man $i$ even if $F_{i,\alpha}$ was very small, it's reasonable to say that the expected energy of men $i$ is bigger than average. Then the possibility of woman $\alpha$ and man $i$ being a $BP$ is smaller than $G_{\alpha,x^{-1}(\alpha)}$. Therefore, the possibility $P_0(\varepsilon)$ that man $i$ with energy $\varepsilon$ cannot find a blocking pair:
\begin{equation}\label{p2}
\begin{split}
P_{0}(\varepsilon)& >\prod_{i=1}^{n\varepsilon}(1-G_{\alpha_{i},x^{-1}(\alpha_{i})})\\
                  &\approx e^{-\sum_{i=1}^{n\varepsilon}G_{\alpha_{i},x^{-1}(\alpha_{i})}}\\
                  &\approx e^{-0.808\sqrt{n}\varepsilon}.
\end{split}
\end{equation}
Hence, the possibility $P_0$ that a man couldn't find a woman to build up a $BP$ can be estimated:

\begin{equation}\label{p3}
\begin{split}
P_{0}&=\int_{0}^{1}f(\varepsilon)P_{0}(\varepsilon)\,d\varepsilon \\
&\approx \int_{0}^{1}f(\varepsilon)e^{-0.808\sqrt{n}\varepsilon}\,d\varepsilon \\
&>\int_{0}^{1}f(\varepsilon)(1-0.808\sqrt{n}\varepsilon)\,d\varepsilon\\
&=1-0.808^{2}=0.35, \\
\end{split}
\end{equation}
which happens to be same with Zhang et al~\cite{scaling}. Furthermore, if we substitute~\eref{fit2} into~\eref{p3},
\begin{equation}\label{p4}
\begin{split}
P_{0}&=\int_{0}^{1}e^{-0.19(\sqrt{n}\varepsilon)^{2}-0.68(\sqrt{n}\varepsilon)+0.02} e^{-0.808\sqrt{n}} \, d\varepsilon\\
     & \approx\int_{0}^{\infty}e^{-0.19\varepsilon^{2}-1.488\varepsilon+0.02}\, d\varepsilon \\
      & =0.60. \\
\end{split}
\end{equation}
The result from numerical simulation is $P_0=0.758$.

It is shown in the numerical simulation that in a system consists of $n$ men and $n$ women, in which there are $0.758n$ men who has no potential mate to build up a $BP$, and $0.242n$ men who have one or more women forming $BPs$ with them. In total, there are $0.325n$ $BPs$ in the system(Fig.5).
\begin{figure}[htb]
\center\scalebox{0.4}[0.4]{\rotatebox{0}{\includegraphics{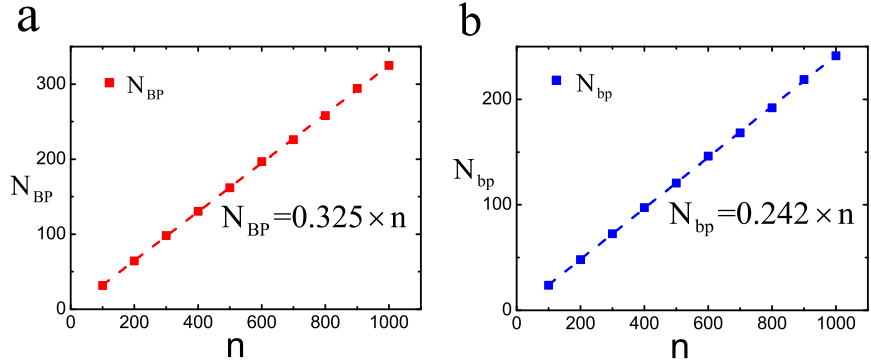}}}
\caption{(a)The amount of $BPs$ $N_{BP}$(red squares) and (b)the number of men who form $BPs$ $N_{bp}$ (blue squares) versus the system scale $n$ in the simulation result. The corresponding dashed lines are the linear fits of the data, slop shown in the figure.}\label{Fig4}
\end{figure}

we can think over the probability that more than one woman who form $BPs$ with man $i$: $m_i$ has a possibility of $75.8\%$ to find no one to form a $BP$ with him, possibility of $24.2\%$ to have one or more women to build up $BPs$. In the $24.2\%$ of men who is one side of the $BPs$, if $m_i$  and $w_\alpha$  forms a $BP$, the other $n-1$ women besides $w_\alpha$  will have a possibility of $75.8\%$ that could not form a $BP$ with man. That is to say, the possibility that there is one and only one woman who builds up a $BP$ with man is $0.242\times0.758$; the possibility that there are two or more women who form $BPs$ with man is $0.242\times0.242$. Among these women, there would be $0.758\times0.242^2$ n women who forms two BPs. And so on we can see the possibility that a man could find exact $k$ women to form $k$ $BPs$ is that:

\begin{equation}\label{p5}
\begin{split}
P_{k}=0.758\times(1-0.758)^{k}, k=0,1,2,\ldots.
\end{split}
\end{equation}
Our simulation result is shown in Fig.6. To further ensure with \eref{p5}, we calculate the total number of BPs:
\begin{equation}\label{bp}
\begin{split}
N_{B.P.}=\sum_{k=1}^{\infty}0.758\times0.242^{k}\times k \times n=0.319n,
\end{split}
\end{equation}
which is in good consistency with former numerical simulation. To each man, averagely, there would be $0.325$ woman who can form a $BP$ with him.

\begin{figure}[htb]
\center\scalebox{0.3}[0.3]{\rotatebox{0}{\includegraphics{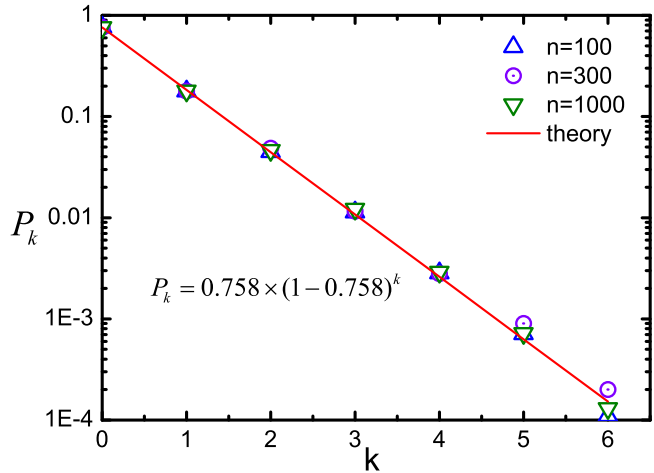}}}
\caption{The probability of $k$ the number of $BPs$ one has in the systems consist of $100$(blue triangles), $300$(purple circles), and $1000$(green inverted triangles) men or women. The red line is the theoretical prediction by previous analysis.}\label{Fig5}
\end{figure}

In order to get a clearer view of the fore mentioned stability issue, here we introduce a parameter which is crucial to the stability of system. Imagine a man $i$ who only acquaint only some of the women, for example, a tiny fraction $k(0<k\ll1)$. In this case, this man $i$ is satisfied only if he couldn't find a $BP$ partner. The possibility which he is satisfied can be roughly estimated as:

\begin{equation}\label{stable1}
\begin{split}
P_{s}=(1-\frac{0.325}{n})^{kn}\approx 1-0.325k \to 1.
\end{split}
\end{equation}
So the possibility of the whole system being stable is that:
\begin{equation}\label{stable2}
\begin{split}
P_{s}^{n}\approx \frac{1}{e}^{0.325nk}\approx 0.72^{K},
\end{split}
\end{equation}
where $K=nk$ is the average number of women this man acquainted.

Here we can see, this parameter $k$ is somehow the key to the stability. As mentioned before~\cite{happier}, if everybody obeyed the allocation of the matchmaker who controls the whole system, this ground state would be the perfect choice only considering about the global circumstances. But in fact, there are some people of them who do not fully focus on the bigger picture (which is very common), then it would be quite possible that someone would like to choose their blocking pair partner rather than the assigned one, should be. If so, Considering in the situation that there was no extra constrains like religion, moral standards and cultural influences, the ground state matching would be broken with several consecutive actions of the pursuit of personal happiness. Generally speaking, as we can see from~\eref{stable2}, if the marriage society is better connected, which means the people will have more relationships with heterosexual agents, we can tell that it's even harder for the system to get stable.~\cite{bachelor}

Is there any way leading us to a stable state of the matching problem? Yes. Roth et al~\cite{path} proved the existence of such a path. They proved that even from arbitrary state, if we keep diminishing the $BPs$, we will never be trapped in a loop, and so we can always find a path to the stable state anyway. But they didn't give out the practical method to find such a path, it's very interesting to study how we can get stable state from the ground state with this method.

\section{Conclusion}
In this paper we studied numerical simulation result of the ground state of Gale-Shapley model with K-M algorithm. With a relaxation on the restrictions of initial assigned energy distribution, the simulation result confirmed the theoretical analysis of the ground state using replica method. The unsolved question about how many blocking pairs are there in the system was also shown in the simulations. Furthermore, we discussed the factors which affects the stability of the system, and found out that the connectivity of the system is crucial to the stability of the ground state.

In the future, the detailed route path from the erratic ground state, to a certain stable state would be of great interests for us to finish. What is the micro dynamics of the system when these hopping agents tried to improve their own situation and  disregarded the cost this bring to the whole system?  After that, it is also thought-provoking to consider the reality meaning of the ground state and the optimal stable solution. May these understanding could better ourselves.

\section*{Acknowledgements}
The authors would like to thank Zhuo-Ming Ren for his valuable comments on this work. This work was partially supported by the EU FP7 Grant 611272 (project GROWTHCOM) and by the Swiss National Science Foundation (grant no.~200020-156188).

\end{document}